\documentclass[aps,10pt,twocolumn,superscriptaddress]{revtex4-1}

\usepackage{amssymb,amsmath,amsthm,amsfonts}
\usepackage[english]{babel}
\usepackage{braket}
\usepackage{graphicx}
\usepackage{tikz}
\usepackage[caption=false]{subfig}
\usepackage[colorlinks]{hyperref}
\usepackage{csquotes}
\usepackage{xifthen}

\usepackage{mathtools}
\mathtoolsset{centercolon}

\colorlet{DarkGreen}{green!70!black}
\colorlet{MarkedOrange}{orange!70!red}
\tikzset{blackdot/.style={circle,inner sep=1.3pt,fill}}
\tikzset{greydot/.style={circle,inner sep=1.3pt,fill,black!60!white}}
\tikzset{reddot/.style={circle,inner sep=1.3pt,fill,color=red}}
\tikzset{bluedot/.style={circle,inner sep=1.3pt,fill,color=blue}}
\tikzset{yellowdot/.style={circle,inner sep=1.3pt,fill,color=yellow}}
\tikzset{greendot/.style={circle,inner sep=1.3pt,fill,color=DarkGreen}}
\tikzset{dottedcell/.style={dotted, fill=black!20!white, fill opacity=0.5}}
\tikzset{largeorangedot/.style={circle,inner sep=1.6pt,fill,color=MarkedOrange}}

\MakeOuterQuote{"}

\theoremstyle{definition}
\newtheorem{example}{Example}
\newtheorem{icase}{Case}

\renewcommand{\d}{\ensuremath{\operatorname{d}\!}}
\DeclareMathOperator{\Tr}{Tr}
\DeclareMathOperator{\dd}{d}
\DeclareMathOperator{\vspan}{span}
\DeclareMathOperator{\poly}{poly}
\newcommand{\Span}[1]{\vspan\left\{#1\right\}}
\newcommand{\HH}{\mathcal H}
\newcommand{\E}{\mathcal E}
\newcommand{\Horacle}{ H_{\text{oracle}}}

\begin{document}

\title{Spatial search by continuous-time quantum walks on crystal lattices}
\author{Andrew M. Childs}
\email{amchilds@uwaterloo.ca}
\affiliation{Department of Combinatorics \& Optimization
             and Institute for Quantum Computing,
             University of Waterloo,
             Waterloo, Ontario, Canada
            }
\author{Yimin Ge}
\email{yge@perimeterinstitute.ca}
\affiliation{Perimeter Institute for Theoretical Physics, Waterloo, Ontario, Canada}

\begin{abstract}
  We consider the problem of searching a general $d$-dimensional lattice of $N$ vertices for a single marked item using a continuous-time quantum walk. We demand locality, but allow the walk to vary periodically on a small scale. By constructing lattice Hamiltonians exhibiting Dirac points in their dispersion relations and exploiting the linear behaviour near a Dirac point, we develop algorithms that solve the problem in a time of $O(\sqrt N)$ for $d>2$ and $O(\sqrt N \log N)$ in $d=2$. In particular, we show that such algorithms exist even for hypercubic lattices in any dimension. Unlike previous continuous-time quantum walk algorithms on hypercubic lattices in low dimensions, our approach does not use external memory.
\end{abstract}
\maketitle

\section{Introduction}

A basic application of quantum computation is solving the problem of finding a marked item among $N$ items. A classical computer takes $\Theta(N)$ steps to find this item with constant probability, but Grover's algorithm \cite{Gro97} shows that a quantum computer can solve this problem using only $O(\sqrt N)$ steps, which is optimal \cite{BBBV97}.

However, Grover's algorithm is unsuited to searching physical databases as it requires performing a reflection about a superposition of all possible items. If the items are distributed in space then this reflection is a nonlocal operation. A locally realisable search algorithm requires that the items are distributed in a $d$-dimensional space and that the quantum computer (viewed as a "quantum robot" \cite{Ben02}) can only perform local operations to explore this database. Aaronson and Ambainis \cite{AA05} constructed such an algorithm that finds a marked item in the optimal time of $O(\sqrt N)$ in $d>2$ dimensions and $O(\sqrt N \poly(\log N))$ in $d=2$. Their algorithm uses a carefully optimised recursive search on subcubes, which raises the question of whether simpler algorithms with the same running time (or better in $d=2$) can be constructed.

Quantum walks on lattices provide  a natural framework for the spatial search problem. Given an $N$-vertex graph $G$, a continuous-time quantum walk is governed by a Hamiltonian $H$ acting on the $N$-dimensional Hilbert space spanned by the states $\ket v$ for all vertices $v$ of $G$. A general state $\ket{\psi(t)}$ is described by $N$ complex amplitudes $\psi_v(t) = \braket{v|\psi(t)}$ and evolves according to the Schr\"odinger equation
\begin{equation}
    i\frac{\d \psi_v(t)}{\d t}= \sum_w H_{vw} \psi_w(t).
\end{equation}
For the spatial search problem, we start in a state $\ket s$ that is independent of the marked item and easy to construct (e.g., the uniform superposition of all vertices) and evolve $\ket s$ for a prescribed time, after which we measure the state in the vertex basis.
The algorithm is successful if the result of the measurement can be used to guess the marked item with constant probability (or with a sufficiently large probability that can be amplified with reasonable computational overhead).
We require that $H$ is \emph{local} in the sense that $H_{vw}$ is nonzero only if $v$ and $w$ are adjacent in $G$.

Following previous continuous-time quantum walk algorithms for spatial search, we  choose $H$ to be of the form
\begin{equation}
    H=H_0+\Horacle,
\end{equation}
where $H_0$ is the \emph{lattice Hamiltonian}, which is independent of the marked item, and $\Horacle$ is the \emph{oracle Hamiltonian}, which perturbs the lattice Hamiltonian to single out the marked item. We require that $H_0$ is local and that the support of $\Horacle$ is localised within a constant radius around the marked item.

Quantum walk algorithms for spatial search have been studied previously. In \cite{CG03}, Childs and Goldstone considered the case where $G$ is a hypercubic lattice in $d$ dimensions and $H_0$ is its adjacency matrix (or equivalently, its Laplacian matrix). It was found that the full quantum speedup of $O(\sqrt N)$ could be achieved in $d>4$ dimensions, whilst for the "critical" dimension $d=4$, a time of $O(\sqrt N\log N)$ is required for a constant probability of success. However, for $d<4$, the algorithm does not provide quadratic speedup over classical algorithms.
Subsequently, Ambainis, Kempe, and Rivosh  \cite{AKR04} found a discrete-time quantum walk algorithm that runs in time $O(\sqrt N)$ for $d>2$ and $O(\sqrt N\log N)$ for $d=2$. Unlike the continuous-time case, a discrete-time quantum walk cannot be defined on the state space of the graph alone but instead requires a coupling to additional degrees of freedom usually called "coins."
Following \cite{AKR04}, Childs and Goldstone \cite{CG04} developed a continuous-time quantum walk algorithm with similar coin registers that has the same performance. In the analysis of \cite{CG03}, the failure of the algorithm in $d<4$ can be viewed as a consequence of a quadratic dispersion relation near the ground state of $H_0$, which is the starting state of the algorithm. Inspired by the Dirac equation, additional "spin" degrees of freedom were introduced as coin registers to construct a Hamiltonian with a "Dirac point" in the dispersion relation. The linear behaviour of the dispersion relation near this point was exploited to reduce the critical dimension from $d=4$ to $d=2$.
Recently, Foulger, Gnutzmann, and Tanner \cite{FGT13} noted that the similar dispersion relation found in the adjacency matrix of a honeycomb lattice can be used to construct a continuous-time quantum walk algorithm with running time $O(\sqrt N\log N)$ in two dimensions without a coin degree of freedom.

In this paper we construct Hamiltonians for efficient spatial search algorithms on hypercubic lattices in $d\geq 2$ dimensions that do not use external memory.
We do this by introducing periodic inhomogeneities to the lattice Hamiltonian $H_0$ instead of taking the adjacency matrix of the graph (which is homogenous across the lattice). This can be naturally treated as a crystal consisting of a periodic lattice with multiple items at each lattice site (sometimes called a lattice with a basis).
The periodic inhomogeneities enable us to construct Hamiltonians with Dirac points, which in turn allows us to reduce the critical dimension from $d=4$ in \cite{CG03} to $d=2$.

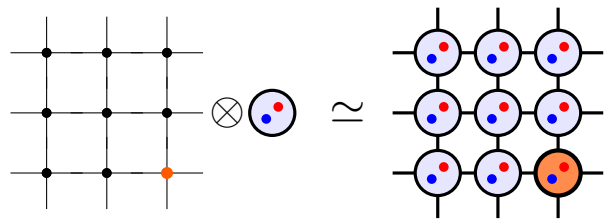
\begin{figure}[h]
    \centering
    \begin{tikzpicture}[scale=0.4]
        \foreach \i in {-4,-2,0}
        {
            \foreach \j in {-1,0,1}
            {
                \draw[blackdot] (\i-1.2,2*\j)
                            -- (\i,2*\j) node [blackdot] {}
                            -- (\i+1.2,2*\j)
                           ;
                \draw[blackdot] (\i,2*\j-1.2)
                            -- (\i,2*\j) node [blackdot] {}
                            -- (\i,2*\j+1.2)
                           ;
            }
        }

        \node () at (2,0) {\LARGE $\otimes$};
        \node[draw, circle, very thick, fill=blue!10, minimum width=6mm] () at (3.5,0) {};
        \node[reddot] () at (3.7,0.2) {};
        \node[bluedot] () at (3.3,-0.2) {};
        \node () at (6,0) {\LARGE $\simeq$};
        \foreach \i in {-1,0,1}
        {
                \draw[very thick] (7.5,2*\i) -- (14.5,2*\i);
                \draw[very thick] (11+2*\i,3.5) -- (11+2*\i,-3.5);
        }
        \foreach \i in {-1,0,1}
        {
            \foreach \j in {-1,0,1}
            {
		\ifthenelse{\cnttest{\i}={1} \AND \cnttest{\j}={-1} }
		{
		  \node[draw, circle, ultra thick, fill=MarkedOrange!70, minimum width=6mm] (\i,\j) at (11+2*\i,2*\j) {};
		}
		{
		  \node[draw, circle, very thick, fill=blue!10, minimum width=6mm] (\i,\j) at (11+2*\i,2*\j) {};
		}
                \node[reddot] () at (11+2*\i+0.2,2*\j+0.2) {};
                \node[bluedot] () at (11+2*\i-0.2,2*\j-0.2) {};
            }
        }
        \draw (0,-2) node[largeorangedot] {};
    \end{tikzpicture}
    \caption{Schematic representation of embedding coin degrees of freedom into the lattice as additional vertices. The resulting new lattice, which in general will not be isomorphic to a simple hypercubic lattice, is a crystal of cells, each containing a number of vertices equal to the dimension of the coin. A naive embedding of the oracle Hamiltonian turns the marked item into an entire marked cell.}
    \label{fig:coin-cell}
\end{figure}

More generally, we present a framework for describing spatial search algorithms using continuous-time quantum walks on arbitrary crystal lattices (subject to certain technical conditions). This construction naturally generalises the results of \cite{FGT13} and is closely related to the ones described in \cite{CG04}. The basic idea, similar to the  staggered fermion formalism \cite{Sus77}, is that coin degrees of freedom can be embedded into the lattice as additional vertices, where the coin registers become cells in the crystal and each cell contains a number of vertices equal to the dimension of the coin space (see Fig.~\ref{fig:coin-cell}). Of course, a naive implementation of this embedding does not result in a hypercubic lattice since the interactions of the coin registers and the original lattice introduce additional edges in the graph, and furthermore turn the marked item into an entire marked cell rather than a single marked vertex. Nevertheless, we show that with further modifications, the structure of a hypercubic lattice can be recovered. The  items within a cell can be viewed as an effective coin, but by a careful choice of oracle Hamiltonian, our approach allows for any vertex to be a possible marked item. As such, the degrees of freedom in our approach correspond directly to the items in the database, unlike in \cite{CG04} where the coin is represented using external memory.

Similar algorithms without additional memory have been proposed and studied numerically for both continuous- and discrete-time quantum walks \cite{PRR05,PR10,PRR10,Fal13}. In \cite{APN13}, Ambainis, Portugal, and Nahimov rigorously analyse the behaviour of the discrete-time algorithm proposed in \cite{Fal13}, which uses a "staggered" quantum walk  consisting of different unitaries at even and odd time steps obtained by different tesselations of the lattice, and obtain the same complexity of $O(\sqrt N \log N)$ for a two-dimensional search.

The remainder of the paper is organised as follows. In Section~\ref{sec:grid} we achieve the same performance as in \cite{CG04}  on a low-dimensional  hypercubic lattice without coin registers by constructing Hamiltonians exhibiting Dirac points.
We then develop a framework for searches on arbitrary crystal lattices in Section~\ref{sec:crystals}, generalising the algorithm  on the hypercubic lattice from Section~\ref{sec:grid} and the algorithm on the honeycomb lattice found in \cite{FGT13}. In Section~\ref{sec:examples} we present several examples of crystal lattices on which efficient search algorithms can be performed.  Finally, we conclude in Section~\ref{sec:discussion} with a brief discussion of the results and some open questions.

\section{Search on the $d$-dimensional hypercubic lattice} \label{sec:grid}

In this section we consider searching a $d$-dimensional hypercubic lattice of $N$ vertices. We construct an algorithm that finds the marked item in time $O(\sqrt N)$   with constant probability for $d>2$ and time $O(\sqrt{N\log N})$ with probability $\Omega(1/\sqrt{\log N})$ for $d=2$. In the latter case, amplitude amplification \cite{BHMT00} can be used to find the marked item with constant probability in time $O(\sqrt N\log N)$.

\subsection{Search Hamiltonian}

We label the $N=L^d$ vertices of a $d$-dimensional hypercubic lattice by $v\in \left[ L\right]^d$, where $\left[m\right] := \{1,\ldots,m\}$. The Hilbert space of the quantum walk is
\begin{equation}
    \HH := \Span{ \ket v \colon v\in \left[ L\right]^d }.
\end{equation}
On this space, consider the lattice Hamiltonian $H_0$ with
\begin{equation} \label{eq:grid-H0}
    H_0\ket{v} = \sum_{i=1}^d (-1)^{v_1+\cdots+v_i}\left(\ket{v+e_i}-\ket{v-e_i}\right),
\end{equation}
where $e_i$ is the unit vector in the $i$th direction.

\begin{figure}[h]
    \centering
    \begin{tikzpicture}[scale=0.5]
     \clip (-2.5,-2.5) rectangle (3.5,3.5);
     \clip (-6,-6) rectangle (6,6);
     \foreach \i in {-3,-2,...,2}
      {
        \foreach \j in {-2,-1,...,2}
        {
         \filldraw[dottedcell] (2*\i-0.3,2*\j-0.3) rectangle (2*\i+1.3,2*\j+1.3);
          \draw (2*\i,2*\j) node [reddot] {}
          -- (2*\i+1,2*\j) node [bluedot] {}
          -- (2*\i+1,2*\j+1) node [greendot] {}
          -- (2*\i,2*\j+1) node [yellowdot] {}
          -- (2*\i,2*\j) node [reddot] {}
          -- (2*\i-1,2*\j) node [bluedot] {}
          -- (2*\i-1,2*\j-1) node [greendot] {}
          -- (2*\i,2*\j-1) node [yellowdot] {}
          -- (2*\i,2*\j) node [reddot] {}
          ;
        }
      }
    \end{tikzpicture}
    \caption{Dividing the hypercubic lattice into $n$ hypercubes (cells), each with $2^d$ vertices.}
    \label{fig:grid-crystal}
\end{figure}
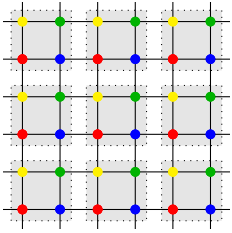

We take $L$ even and impose periodic boundary conditions, so that this Hamiltonian is invariant under translations  of length $2$, that is, $H_0$ commutes with the translation operators $T_i$ defined by
\begin{equation}\label{eq:grid-translation}
    T_i\ket v = \ket{v+2e_i}.
\end{equation}
It is therefore convenient to consider the lattice as a crystal consisting of $n:=N/2^d$ cells, each a $d$-dimensional hypercube with $2^d$ vertices (see Fig.~\ref{fig:grid-crystal}). We define $l:=L/2$.

\begin{figure}[h]
  \centering
  \includegraphics[width=0.4\textwidth]{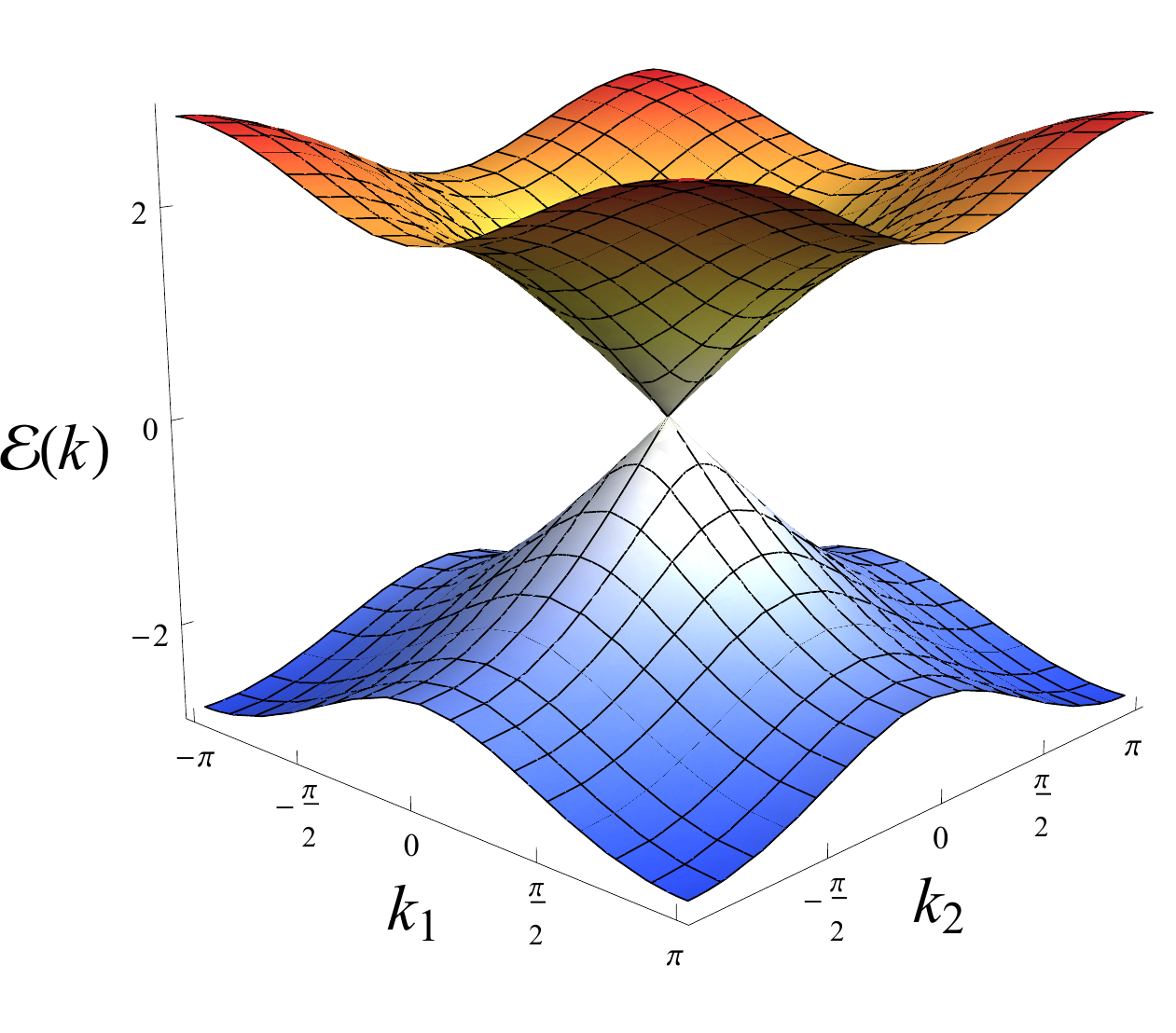}
  \caption{Dispersion relation for $d=2$.}
  \label{fig:grid-dispersion}
\end{figure}

We can thus write
\begin{equation}
    \ket v = \ket{x,\sigma},
\end{equation}
where $x\in \left[l\right]^d$ labels the cell and $\sigma\in \mathbb Z_2^d$ labels the vertex within the cell, with
\begin{align}
    x_i&=\left\lfloor \frac{v_i}2\right\rfloor,\\
    \sigma &= v-2x.
\end{align}
Writing $\overline{\sigma_i}:= 1-\sigma_i$ to denote the logical negation of the $i$th component of $\sigma$, the lattice Hamiltonian acts as
\begin{align}
    H_0\ket{x,\sigma} &= \sum_{i=1}^d \begin{aligned}[t] (-1)^{s_i(\sigma)}(&\ket{x+\sigma_ie_i,\sigma+e_i} \\
    -&\ket{x-\overline{\sigma_i}e_i,\sigma+e_i}),\end{aligned}  \label{eq:grid-H0-position}
\end{align}
where $s_i(\sigma):=\sigma_1+\cdots+\sigma_i$.
Translational invariance by \eqref{eq:grid-translation} implies that $H_0$ is block-diagonal of block size $2^d$ in the Fourier basis given by
\begin{align}\label{eq:grid-fourier}
    \ket{k,\sigma} &:= \frac1{\sqrt n} \sum_{x\in \left[l\right]^d} e^{ik\cdot x} \ket{x,\sigma},\\
    k_i&=\frac{2\pi m_j}{l},\qquad m_j\in\left[l\right].
\end{align}
In particular,
\begin{align}
    &H_0\ket{k,\sigma} \nonumber \\
    &= \sum_{i=1}^d (-1)^{s_i(\sigma)}  \left( e^{-ik_i\sigma_i } - e^{ik_i\overline{\sigma_i}}\right)\ket{k,\sigma+e_i} \label{eq:grid-fourier-coeff1} \\
        &= \sum_{i=1}^d (-1)^{s_i(\sigma)}  \big( (-1)^{\sigma_i} (1-\cos k_i)  -i\sin k_i \big) \ket{k,\sigma+e_i}. \label{eq:grid-fourier-coeff2}
\end{align}
We can thus write
\begin{equation} \label{eq:grid-directsum}
    \HH \simeq \bigoplus_k \HH_k
\end{equation}
and
\begin{equation}
    H_0 = \sum_k H_0(k), \label{eq:grid-block1}
\end{equation}
where $\HH_k:=\Span{\ket{k,\sigma} : \sigma\in\mathbb Z_2^d}$ and each $H_0(k)$ acts only on $\HH_k$.
To find the eigenvalues of $H_0$, notice that
\begin{equation}
    H_0(k)^2\ket{k,\sigma} = \E(k)^2\ket{k,\sigma},
\end{equation}
where
\begin{equation}\label{eq:grid-dispersion}
    \E(k): = \sqrt{ \sum_{i=1}^d \left(\sin^2 k_i + (1-\cos k_i)^2\right)}.
\end{equation}
Thus the eigenvalues of $H_0(k)$ are $\pm\E(k)$ and because $\Tr H_0(k)=0$ (indeed, $\bra{k,\sigma}H_0\ket{k,\sigma}=0$ for all $k,\sigma$), both eigenvalues have multiplicity $2^{d-1}$.
Notice that $k=0$ is the unique value of $k$ for which $\E(k)=0$ and that near $k=0$ the dispersion relation \eqref{eq:grid-dispersion} behaves linearly: $\E(k) \approx \lvert k\rvert$ for small values of $k$ (see Fig.~\ref{fig:grid-dispersion}).

The full algorithm is as follows. Suppose $\ket{w,\alpha}$ is the marked item (where $w\in\left[l\right]^d$ labels the hypercube and $\alpha\in\mathbb Z_2^d$ the vertex within the hypercube). We begin in the uniform superposition of all items with $\sigma=\alpha$,
\begin{equation} \label{eq:grid-starting-state}
    \ket s := \frac 1{\sqrt n} \sum_{x} \ket{x,\alpha},
\end{equation}
and evolve with the Hamiltonian
\begin{equation}
    H := H_0 + \Horacle
\end{equation}
for some time $T$, where
\begin{equation} \label{eq:grid-oracle}
    \Horacle := - \ket{w,\alpha}\bra{w,\alpha} H_0 - H_0 \ket{w,\alpha}\bra{w,\alpha}
\end{equation}
is the oracle Hamiltonian, generalising the expression chosen in \cite{FGT13}. Notice that this choice differs from the naive choice of $\Horacle \propto \ket{w,\alpha}\bra{w,\alpha}$ used in \cite{CG03}. This modification accounts for the symmetry of the dispersion relation \eqref{eq:grid-dispersion} and the fact that the graph is $2^d$-partite in the site label $\sigma\in\mathbb Z_2^d$ (we discuss this choice further in Section~\ref{sec:crystals}).

\begin{figure}[h]
  \centering
  \subfloat[][$d=2$]{
    \includegraphics[width=.25\textwidth,clip,trim=0 0 0 0]{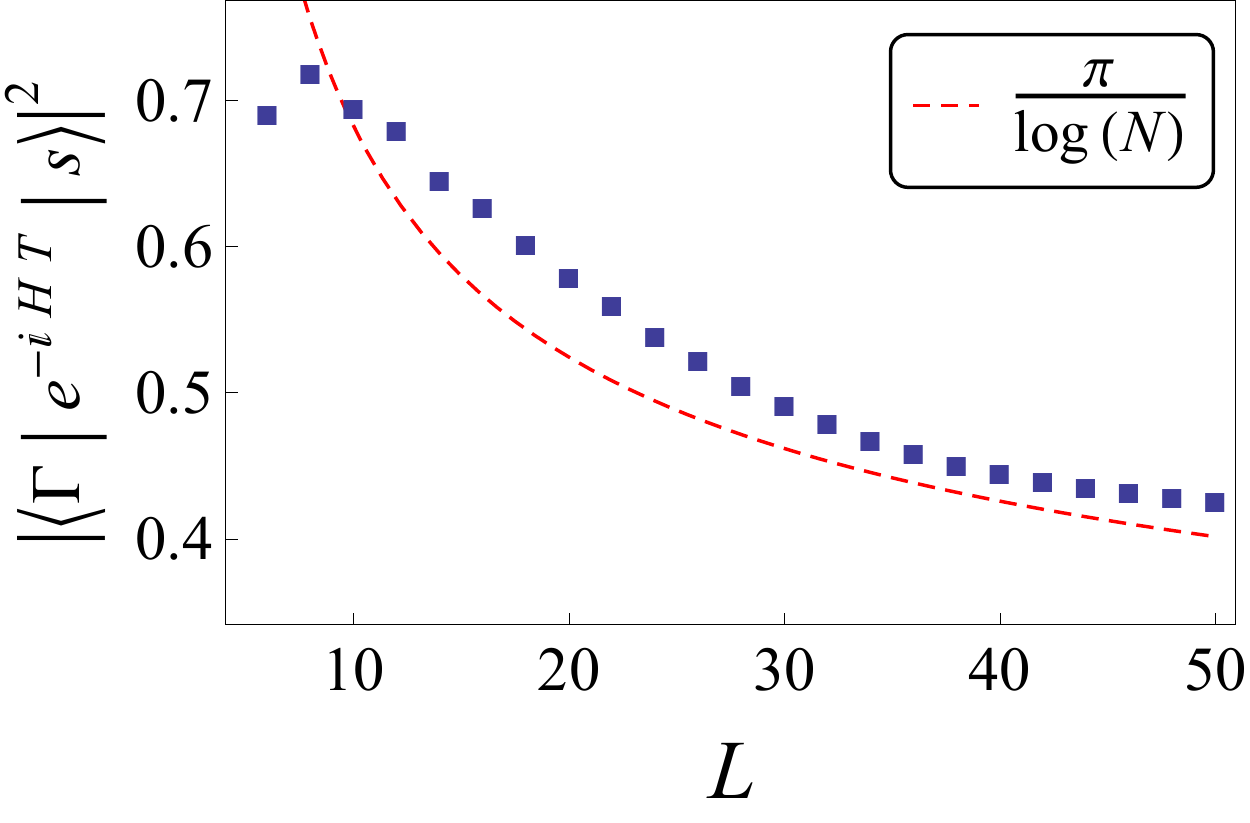}
    \label{fig:numericL-2D}}
  \subfloat[][$d=3$]{
    \includegraphics[width=.21\textwidth,clip,trim=0 1mm 0 0]{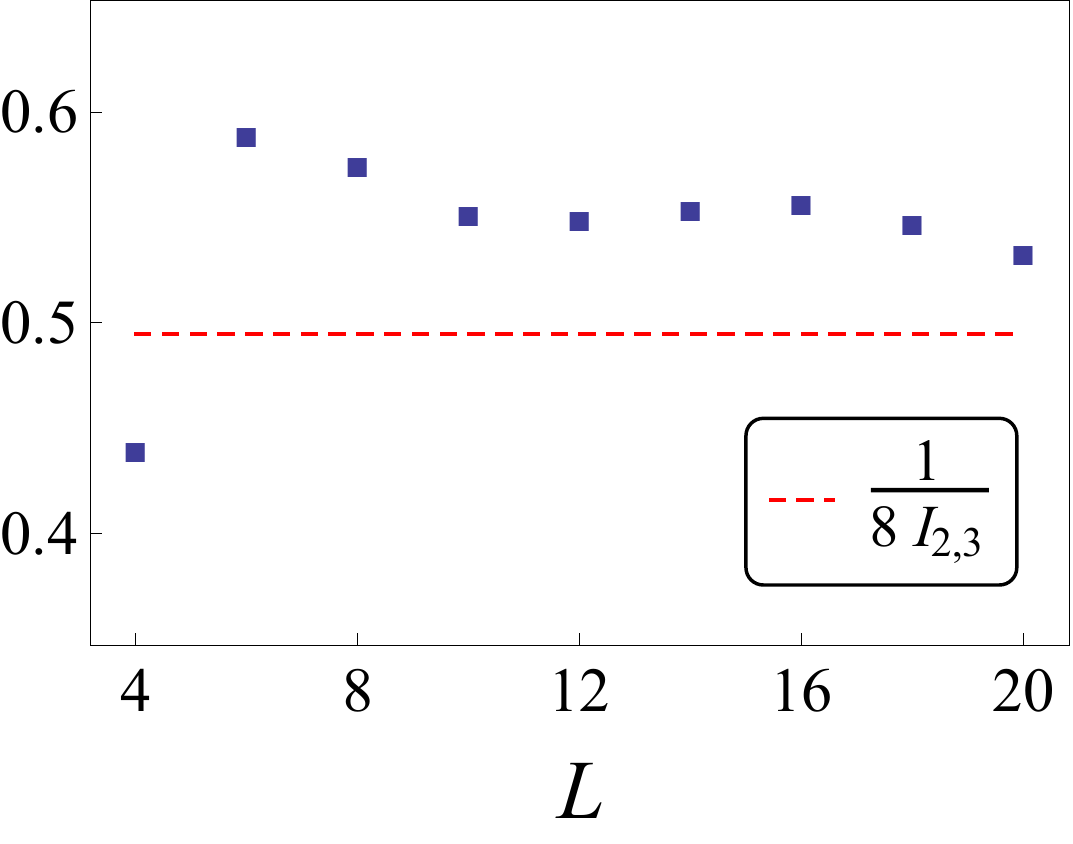}
    \label{fig:numericL-3D}}
   \caption{Numerical values of $\lvert\bra\Gamma e^{-iHT}\ket{s}\rvert ^2$ for increasing lattice sizes. (a) In $d=2$, the squared overlap at $T=\sqrt{\frac{\pi}{64}N\log N}$ is $\Omega(1/{\log N})$. (b) In $d=3$, the squared overlap at $T=\frac\pi2\sqrt{ I_{2,3}N}$ is approximately $1/8I_{2,3}$, where $I_{2,3}$ is a constant defined below.}
   \label{fig:numericL}
\end{figure}

\begin{figure}[h]
  \centering
    \includegraphics[width=0.4\textwidth]{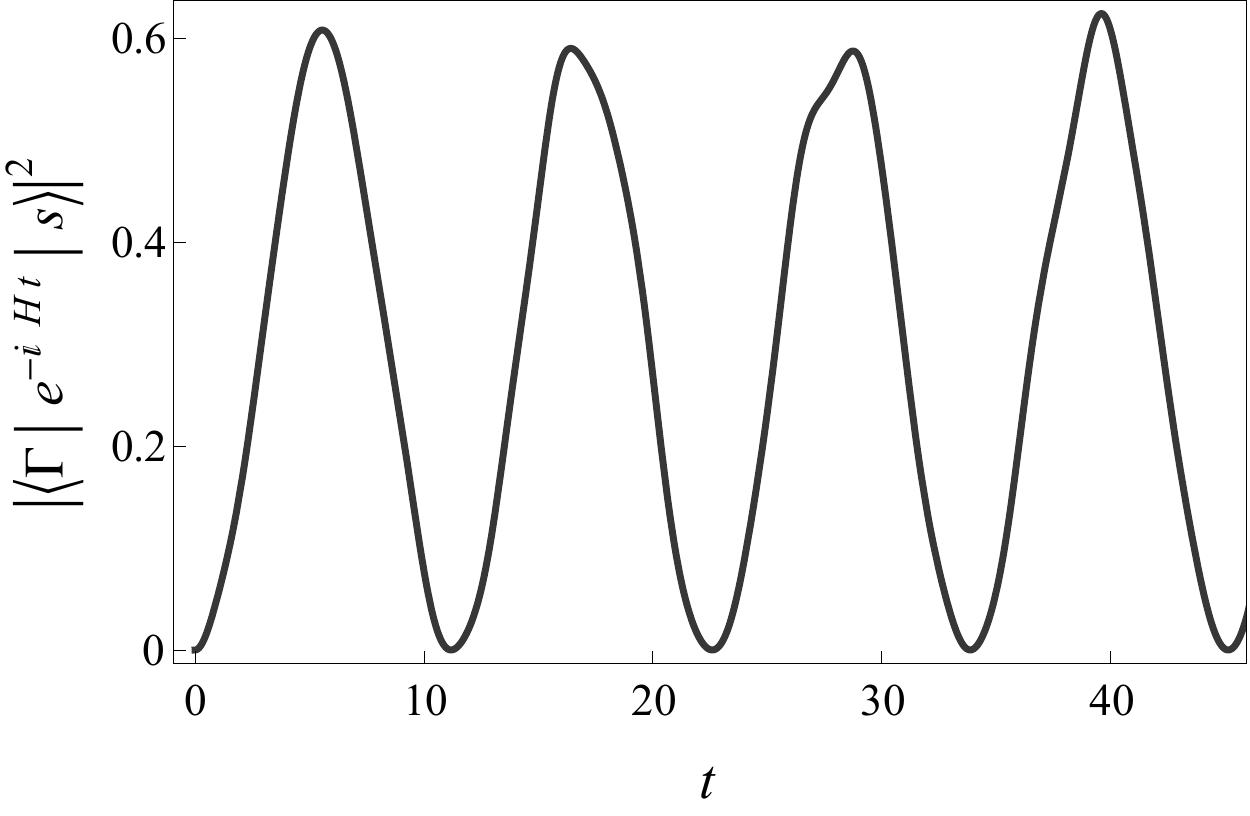}
    \caption{Time-dependent squared overlap $\lvert\bra\Gamma e^{-iHt}\ket{s}\rvert ^2$ for an $8 \times 8 \times 8$ cubic lattice.}
       \label{fig:numerics-T}
\end{figure}

We will show that for $d\geq 3$, the evolved state $e^{-iHT}\ket s$ has constant overlap with the normalised state
\begin{equation}\label{eq:grid-gamma}
    \ket\Gamma := \frac{1}{\sqrt{2^d}}H_0\ket{w,\alpha}
\end{equation}
after time $T=O(\sqrt N)$ (see Fig.~\ref{fig:numericL-3D} for a numerical example).
Since $\ket\Gamma$ only has nonzero amplitudes on the neighbours of $\ket{w,\alpha}$, we thus find a neighbour of $\ket{w,\alpha}$ with constant probability, which in turn lets us guess $\ket{w,\alpha}$ itself with constant probability of success.
As in previous quantum search algorithms \cite{Gro97,CG03,AKR04,CG04,FGT13}, the success probability oscillates and a measurement should be performed at the correct time to maximise the success probability (see Fig.~\ref{fig:numerics-T}).
For $d=2$, the overlap is $\Omega(1/\sqrt{\log N})$ after time $T=O(\sqrt{N\log N})$ (see Fig.~\ref{fig:numericL-2D}),
so amplitude amplification can be used to obtain constant overlap with $\ket\Gamma$ after time $O(\sqrt N\log N)$. Notice, however, that $\ket s$ depends on $\alpha$, which in turn depends on the unknown marked item. Therefore, we run the algorithm multiple times with different starting states, once for each of the $2^d$ possible values for $\alpha$. For fixed $d$, this increases the overall complexity only by a constant factor.

\subsection{Analysis of the algorithm}\label{sec:grid-analysis}
To analyse the algorithm, we determine the spectrum of $H$ using the spectrum of $H_0$. We use similar techniques as in \cite{CG03,AKR04,CG04,FGT13}.

First, notice that $\bra{w,\alpha}H_0\ket{w,\alpha}=0$ implies
\begin{equation}
    H\ket{w,\alpha}=0,
\end{equation}
i.e., $\ket{w,\alpha}$ is an eigenvector of $H$ with eigenvalue zero.
Let $\ket{\psi_a}$ be an eigenvector of $H$  of eigenvalue $E_a\neq 0$, which we assume to be not in the spectrum of $H_0$. Then,  in particular,
\begin{equation}
    E_a\braket{w,\alpha|\psi_a} = \bra{w,\alpha}H\ket{\psi_a} = 0,
\end{equation}
so
\begin{equation}
    \label{eq:grid-orthogonal}
    \braket{w,\alpha|\psi_a}=0.
\end{equation}
Now, $H\ket{\psi_a}=E_a\ket{\psi_a}$ and \eqref{eq:grid-orthogonal} imply that
\begin{equation}\label{eq:grid-eval-equiv}
    (H_0-E_a)\ket{\psi_a} = \ket{w,\alpha}\bra{w,\alpha}H_0\ket{\psi_a}
\end{equation}
and since we assumed that $E_a$ is not in the spectrum of $H_0$, this implies that
\begin{equation} \label{eq:grid-eval1}
    \ket{\psi_a} = \sqrt{R_a}(H_0-E_a)^{-1}\ket{ w,\alpha},
\end{equation}
where
\begin{equation}\label{eq:grid-Ra-1}
    \sqrt{R_a}:=\bra{w,\alpha} H_0\ket{\psi_a}\neq 0.
\end{equation}
By choice of phase, we can assume without loss of generality that $\sqrt{R_a}>0$. Then \eqref{eq:grid-orthogonal} implies the eigenvalue condition
\begin{equation} \label{eq:grid-condition}
    F(E_a)=0,
\end{equation}
where
\begin{equation}\label{eq:grid-F-def}
    F(E) := \bra{w,\alpha} (H_0-E)^{-1}\ket{w,\alpha}.
\end{equation}
Note that \eqref{eq:grid-condition} differs from the eigenvalue condition obtained in \cite{CG03,CG04}, which was $F(E_a)=1$. This is a direct consequence of the different choice of the oracle Hamiltonian \eqref{eq:grid-oracle}.

So far, we have only shown that \eqref{eq:grid-condition} is a necessary condition for $E_a$ to be an eigenvalue of $H$, but \eqref{eq:grid-condition} is also sufficient for $E_a$ to be an eigenvalue. Indeed, suppose that $E_a\neq 0$ is not contained in the spectrum of $H_0$ and satisfies \eqref{eq:grid-condition}.
The existence of a vector $\ket{\psi_a}$ satisfying $H\ket{\psi_a} = E_a\ket{\psi_a}$ is equivalent to the existence of a vector $\ket{\psi_a}$ satisfying
\begin{align}
    \ket{\psi_a} &= (H_0-E_a)^{-1}\ket{ w,\alpha}\bra{w,\alpha}H_0\ket{\psi_a}  \nonumber \\ \
                &\quad +(H_0-E_a)^{-1} H_0\ket{w,\alpha}\braket{{w,\alpha}|{\psi_a}}.
\end{align}
Equivalently, the operator
\begin{align}
    X(E_a) &:=(H_0-E_a)^{-1}\ket{ w,\alpha}\bra{w,\alpha} H_0 \nonumber \\
        &\quad + (H_0-E_a)^{-1} H_0\ket{w,\alpha}\bra{w,\alpha}\label{eq:grid-sufficient}
\end{align}
has an eigenvalue of $1$.
Since
\begin{equation}
    (H_0-E_a)^{-1} H_0 = 1+(H_0-E_a)^{-1}E_a,
\end{equation}
the assumption \eqref{eq:grid-condition} implies that
\begin{equation}
    \bra{w,\alpha}(H_0-E_a)^{-1} H_0\ket{w,\alpha}=1,
\end{equation}
so that
\begin{equation}
    X(E_a)^\dagger\ket{w,\alpha}=\ket{w,\alpha}.
\end{equation}
But since a finite-dimensional Hermitian operator has the same eigenvalues as its adjoint, $X(E_a)$ also has an eigenvalue $1$.

Furthermore, notice that normalisation of \eqref{eq:grid-eval1} implies that
\begin{equation} \label{eq:grid-Ra-2}
    R_a^{-1} =  \bra{w,\alpha}(H_0-E_a)^{-2}\ket{w,\alpha} = F'(E_a).
\end{equation}
We also need the overlaps of the eigenvectors of $H$ with the starting state. By taking the inner product of \eqref{eq:grid-starting-state} and \eqref{eq:grid-eval1}, we find
\begin{equation} \label{eq:grid-spsi}
    \braket{\psi_a|s} = -\frac{\sqrt{R_a}}{E_a\sqrt{n}} = -\frac{1}{E_a\sqrt{nF'(E_a)}}.
\end{equation}

For $k\neq 0$, let $\HH_k^\pm < \HH_k$ be the eigenspaces of the eigenvalues $\pm \E(k)$, respectively, and let $P_k^\pm$ be the projectors onto $\HH_k^\pm$. Furthermore, let $P_0$ be the projector onto $\HH_0$.
For $k\neq 0$, we have
\begin{equation} \label{eq:grid-H0-sym}
      H_0(k)=\E(k)\left(P_k^+ - P_k^- \right)
\end{equation}
and
\begin{equation} \label{eq:grid-block-complete}
   P_k:= P_k^+ + P_k^- = \sum_{\sigma\in \mathbb Z_2^d}\ket{k,\sigma}\bra{k,\sigma}.
\end{equation}
Equations \eqref{eq:grid-block-complete} and \eqref{eq:grid-fourier} imply that, for all $k$,
\begin{equation} \label{eq:grid-projection-norm}
    \left\| P_k\ket{w,\alpha} \right\|^2 = \sum_{\sigma\in \mathbb Z_2^d}\lvert\braket{w,\alpha|k,\sigma}\rvert^2 = \frac1n.
\end{equation}

Let $\tilde H_0$ be the restriction of $H_0$ to the subspace
\begin{equation}
    \tilde \HH := \bigoplus_{k\neq 0} \HH_k.
\end{equation}
Notice that $\tilde H_0$ is invertible. Let $\ket{w_0} := P_0\ket{w,\alpha}\in \HH_0$ and $\ket{\tilde w} := \ket{w,\alpha} - \ket{w_0}\in\tilde\HH$ be the projections of $\ket{w,\alpha}$ onto $\HH_0$ and $\tilde \HH$, respectively.

Since $H_0\ket{w_0}=0$, we can write \eqref{eq:grid-F-def} as
\begin{align}
    F(E) &= -\left\|\ket{w_0}\right\|^2 \frac{1}{E} + \bra{\tilde w}\left( H_0-E\right)^{-1} \ket{\tilde w} \label{eq:grid-separate-0} \\
        &= -\frac{1}{nE} + \bra{\tilde w}(\tilde H_0-E)^{-1} \ket{\tilde w},\label{eq:grid-separate-0-1}
\end{align}
where the last equality follows from \eqref{eq:grid-projection-norm}.

We now analyse the eigenvalue condition \eqref{eq:grid-condition} by Taylor expansion. We rigorously justify these approximations in Section~\ref{sec:grid-rigorous}.
If $|E| \ll \E(k)$ for all $k\neq 0$, we can Taylor expand the second term in \eqref{eq:grid-separate-0-1} to obtain
\begin{equation}
    \label{eq:grid-taylor1}
    F(E)\approx -\frac{1}{nE} + \bra{\tilde w} \tilde H_0^{-1}\ket{\tilde w} +  E \bra{\tilde w} \tilde H_0^{-2}\ket{\tilde w}.
\end{equation}

The middle term vanishes since $\bra{w,\alpha}H_0(k)\ket{w,\alpha}=0$ for all $k$, so
\begin{align}
    \bra{\tilde w} \tilde H_0^{-1}\ket{\tilde w} &= \sum_{k\neq 0}\frac{1}{\E(k)}\bra{w,\alpha}\left(P_k^+ - P_k^-\right)\ket{w,\alpha} \nonumber \\
    &= \sum_{k\neq 0}\frac{1}{\E(k)^2}\bra{w,\alpha}H_0(k)\ket{w,\alpha} = 0.
\end{align}
In $d>2$ dimensions, using \eqref{eq:grid-projection-norm} we can approximate the last term as
\begin{align}
    \bra{\tilde w} \tilde H_0^{-2}\ket{\tilde w} &= \sum_{k\neq 0}\left\| P_k\ket{w,\alpha} \right\|^2 \frac{1}{\E(k)^2} \\
        &= \frac1n \sum_{k\neq 0} \frac{1}{\E(k)^2} \\
        &\approx \frac{1}{(2\pi)^d}\int_{-\pi}^\pi \frac{\d^d k}{\E(k)^2}  =: I_{2,d} \label{eq:grid-integral}
\end{align}
where the integral converges for $d>2$ (see Table \ref{tab:grid-numerical-I2} for numerical values of $I_{2,d}$).

\begin{table}[h]
    \centering
    \begin{tabular}{c|l}
      $d$ & \multicolumn{1}{c}{$I_{2,d}$} \\
      \hline
      3 & $0.2527$ \\
      4 & $0.1549$ \\
      5 & $0.1156$ \\
      6 & $0.0931$
    \end{tabular}
    \caption{Numerical values for $I_{2,d}$.}
    \label{tab:grid-numerical-I2}
\end{table}

We thus obtain
\begin{equation} \label{eq:grid-F-final}
    F(E)\approx -\frac1{nE} + I_{2,d} E,
\end{equation}
which by \eqref{eq:grid-condition} gives us the eigenvalues
\begin{equation}\label{eq:grid-eigenvalues}
    E_\pm \approx \pm\frac1{\sqrt{nI_{2,d}}}.
\end{equation}
Notice that they indeed satisfy $|E_\pm |\ll \E(k)$ for all $k\neq 0$. It furthermore can be shown (see Section~\ref{sec:grid-rigorous}) that for these values of $E_\pm$, the higher-order terms in \eqref{eq:grid-taylor1} are negligible. Using \eqref{eq:grid-F-final}, we also obtain
\begin{equation} \label{eq:grid-F'}
    F'(E_\pm) \approx 2I_{2,d}.
\end{equation}
Let $\ket{\psi_\pm}$ be the corresponding eigenstates of $H$. Using \eqref{eq:grid-spsi}, we see that $\braket{\psi_\pm|s} \approx \mp \frac{1}{\sqrt2}$, so the starting state is
\begin{equation}
    \ket s \approx \frac{1}{\sqrt 2} (\ket{\psi_-}  - \ket{\psi_+}).
\end{equation}
Evolving for time $T=\pi/(2|E_\pm|)$ gives (up to a global phase) the state
\begin{equation}
    e^{-iHT}\ket s \approx \frac{1}{\sqrt 2} (\ket{\psi_-} + \ket{\psi_+}),
\end{equation}
which by \eqref{eq:grid-Ra-1} and \eqref{eq:grid-Ra-2} has an overlap with $\ket\Gamma$ (defined in \eqref{eq:grid-gamma}) of approximately
\begin{align} \label{eq:grid-overlap}
    \lvert\bra\Gamma e^{-iHT}\ket{s}\rvert &\approx \frac{1}{\sqrt{2^{d+1}}}\left(\frac1{\sqrt{F'(E_-)}}+\frac1{\sqrt{F'(E_+)}}\right) \nonumber \\
    &\approx\frac{1}{\sqrt{2^dI_{2,d}}},
\end{align}
which is constant.

For $d=2$, the integral $I_{2,d}$ diverges logarithmically. Specifically, equations \eqref{eq:grid-F-final}--\eqref{eq:grid-overlap} hold with $I_{2,d}$ replaced with
\begin{equation}
	I_{2,d} = \frac{1}{4\pi} \log N + O(1),
\end{equation}
which can be seen as follows. The smallest nonzero value of $k$ satisfies $|k| = 2\pi/l$. Letting $U:=\left\{k \in[-\pi,\pi]^d \colon |k|\geq 2\pi/l \right\}$, we can approximate the last term of \eqref{eq:grid-taylor1} as
\begin{align}
    \bra{\tilde w} \tilde H_0^{-2}\ket{\tilde w} &= \frac1n \sum_{k\neq 0} \frac{1}{\E(k)^2} \label{eq:grid-2d-I2-1}\\
                &= \frac{1}{(2\pi)^2}\int_{U} \frac{\d^2 k}{\E(k)^2} + O(1) \\
                &= \frac1{2\pi}\int_{\frac{2\pi}{l}}^\pi \frac{\d k}{k} + O(1) \\
                &= \frac{1}{4\pi}\log N + O(1) \label{eq:grid-2d-I2-2}.
\end{align}

Thus we find that evolving for a time $T=O(\sqrt{N\log N})$ produces a state with an overlap of $\Omega(1/\sqrt{\log N})$ on $\ket\Gamma$.

\subsection{Fine-tuning the Hamiltonian} \label{sec:grid-finetuning}

In previous continuous-time quantum walk algorithms for spatial search \cite{CG03,CG04}, the full Hamiltonian was of the form
\begin{equation}\label{eq:grid-H-gamma}
    H = \gamma H_0 +\Horacle,
\end{equation}
where $\gamma$ was an adjustable parameter that had to be fine-tuned to a critical value. In the analysis above, we have already implicitly tuned this parameter to $\gamma=1$, which is the critical value for this algorithm. In practice, this exact fine-tuning might be difficult to achieve. We now briefly consider the effect of varying $\gamma$ away from $1$.

It is easy to verify that if the Hamiltonian is replaced with \eqref{eq:grid-H-gamma}, the eigenvalue condition \eqref{eq:grid-condition} becomes
\begin{equation}
    F(E_a) = \frac{f(\gamma)}{E_a},
\end{equation}
where now
\begin{equation}
    F(E): = \bra{w,\alpha}(\gamma H_0-E)^{-1}\ket{w,\alpha}
\end{equation}
and $f(\gamma) := (\gamma-1)^2 / (2\gamma -1)$. Repeating the analysis of Section~\ref{sec:grid-analysis} results in the eigenvalues
\begin{equation}
    E_\pm = \pm\sqrt{\frac{\gamma^2(1+nf(\gamma))}{nI_2}}.
\end{equation}
For $\gamma$ close to $1$, $f(\gamma) = (\gamma-1)^2 + O\bigl((\gamma-1)^4\bigr)$.
Thus the algorithm behaves similarly provided $\gamma$ can be fine-tuned to a precision of $|\gamma-1| = o(1/\sqrt N)$.

\subsection{Validity of Taylor expansion}\label{sec:grid-rigorous}
We now give a rigorous justification of the approximations used in \eqref{eq:grid-taylor1}--\eqref{eq:grid-F'}. Notice that we only need to justify these steps for $E = \Theta(1/\sqrt n)$ for $d\geq 3$ and $E=\Theta(1/\sqrt{n\log n})$ for $d=2$.

The second term in \eqref{eq:grid-separate-0-1} can be written as
\begin{equation} \label{eq:grid-rigorous-1}
    \bra{\tilde w}(\tilde H_0-E)^{-1} \ket{\tilde w}
        = \frac1n \sum\limits_{\substack {k\neq 0\\ \eta=\pm} }  \frac{1}{\eta\E(k) -E}.
\end{equation}
This is just a sum of $2(n-1)$ geometric series of ratio $\pm\E(k)/E$. The radius of convergence of \eqref{eq:grid-rigorous-1} as a power series in $E$ is thus the smallest $|\E(k)|$, which is $\Theta(n^{-1/d})$. The Taylor expansion \eqref{eq:grid-taylor1} as well as taking the termwise derivative \eqref{eq:grid-F'} are thus justified for the values of $E_\pm$ that lie within the radius of convergence for sufficiently large $n$.

To show that it suffices to expand to first order in $E$, notice that the $m$th coefficient in the Taylor expansion \eqref{eq:grid-taylor1} is
\begin{equation}
    \bra{\tilde w}\tilde H_0^{-m} \ket{\tilde w} = \frac1n\sum\limits_{\substack {k\neq 0\\ \eta=\pm} }  \frac{1}{(\eta\E(k))^{m}}.
\end{equation}
A similar analysis to \eqref{eq:grid-2d-I2-1}--\eqref{eq:grid-2d-I2-2} shows that this is finite for $m<d$, $O(\log n)$
for $m=d$, and at most $cn^{m/d-1}$ for $m>d$, where $c>0$ is a constant independent of $m$.
Thus, for $d\geq 3$ and $E=\Theta(1/\sqrt n)$, we see that the sum of all higher-order terms in \eqref{eq:grid-taylor1} is
\begin{align}
    &O\biggl( n^{-\frac d2}\log n + \sum_{m=d+1}^{\infty} n^{-\frac{m-1}2 + \frac md -1}\biggr) \nonumber \\
    &\quad= O\Bigl(n^{-\frac{(d+1)(d-2)}{2d} - \frac12}\Bigr) \nonumber \\
    &\quad= o(n^{-1/2}).
\end{align}

A similar argument shows that the higher-order terms are $o(1/\sqrt{n\log n})$ when $d=2$ and $E=\Theta(1/\sqrt{n\log n})$.

Finally, we relate the approximate eigenvalues in \eqref{eq:grid-eigenvalues} to the actual values. To do this, let $E_+ := 1/\sqrt{nI_{2,d}}$ be the approximate and $\tilde E$ the true solution of $F(E)=0$ that is closest to $E_+$ (a similar argument also holds for $E_-$). For $d\geq 3$, it suffices to show that $|\tilde E-E_+| = o(n^{-1/2})$.
Direct substitution into \eqref{eq:grid-taylor1} shows that there exist constants $c_1,c_2>0$ independent of $n$ such that, for sufficiently large $n$, $F(c_1 n^{-1/2}) < 0$ and $F(c_2 n^{-1/2})  > 0$. Thus, by the intermediate value theorem, $\tilde E = \Theta(n^{-1/2})$.
By expanding $F$ around $ E_+$, Taylor's theorem shows that
\begin{equation}\label{eq:grid-taylor-justification}
    0=F(\tilde E) = F(E_+) + (\tilde E - E_+)F'(e)
\end{equation}
for some $e$ between $\tilde E$ and $E_+$. Since $e=\Theta(n^{-1/2})$ is within the radius of convergence of the Taylor series, the termwise derivative of $F$ shows that $F'(e)$ is bounded below by a positive constant as $n\rightarrow \infty$. But $F(E_+)$ is just the sum of higher-order terms of the Taylor expansion which, as we have seen, is at most $O(n^{-(d+1)(d-2)/2d-1/2}) = o(n^{-1/2})$. Thus \eqref{eq:grid-taylor-justification} shows that $|\tilde E - E_+| = o(n^{-1/2})$, as required.

A similar analysis shows that for $d=2$, $|\tilde E - E_+| = o(1/\sqrt{n\log n})$.

\section{Search on general crystal lattices} \label{sec:crystals}

The algorithm introduced in the previous section relies  on the behaviour of the dispersion relation \eqref{eq:grid-dispersion} near the energy of the starting state $\ket s$. Specifically, the linear behaviour of the dispersion relation near $k=0$  is responsible for the efficiency of the algorithm even in low dimensions. In \cite{CG03}, the quadratic instead of linear dispersion near the eigenvalue of the starting state implies that $I_{2,d}$ only converges for $d>4$ instead of $d>2$, thus resulting in a search algorithm with quadratic speedup only for $d\geq 4$ instead of $d\geq 2$.

Values of $k$ with linear behaviour in the dispersion relation are commonly referred to as \emph{Dirac points}. For our purposes, we say that a dispersion relation $\E(k)$ has a Dirac point at $k=\tilde k$ if there exist constants $c,K>0$ such that $|\E(\tilde k +\delta ) - \E(\tilde k)| > c|\delta|$ for all $\delta  \in \mathbb R^d$ with $|\delta| < K$.

In this section, we generalise the results from the previous section to any lattice Hamiltonian whose dispersion relation has a finite number of Dirac points of the same energy.

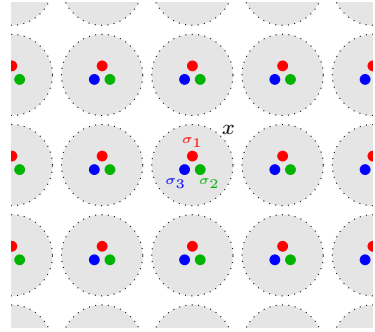
\begin{figure}[h]
  \centering
    \begin{tikzpicture}[scale=0.6]
        \clip (-4,-3.6) rectangle (4,3.6);
        \foreach \i in {-3,-2,...,3}
          {
            \foreach \j in {-3,-2,...,3}
            {
                \filldraw[dottedcell] (2*\i,2*\j) circle (0.9);
                \node[draw,reddot] at (2*\i,2*\j+0.2) {};
                \node[draw,greendot] at (2*\i+0.2*cos{30},2*\j-0.2*sin{30}) {};
                \node[draw,bluedot] at (2*\i-0.2*cos{30},2*\j-0.2*sin{30}) {};
            }
          }
          \node [color=red, above] at (0,0.2) {\tiny$\sigma_1$};
          \node [color=DarkGreen, below] at (0.2*cos{30}+0.2,-0.2*sin{30}) {\tiny$\sigma_2$};
          \node [color=blue, below] at (-0.2*cos{30}-0.2,-0.2*sin{30}) {\tiny$\sigma_3$};
          \node at (0.8,0.8) {\footnotesize$x$};
        \end{tikzpicture}
  \caption{Schematic representation of a crystal with $r=3$.}
  \label{fig:crystal}
\end{figure}

Suppose we have $N=nr$ items (vertices) arranged in a crystal of $n$ cells in a lattice, each of which contains $r$ items (see Fig.~\ref{fig:crystal}).

We can  assume without loss of generality that the underlying lattice is a hypercubic lattice in $d$ dimensions of linear size $l=n^{1/d}$. We impose periodic boundary conditions.
Let $\Sigma$ be a set of labels for the items within a cell, with $|\Sigma|=r$. Then, as before, the items in the crystal are labelled by a pair $(x,\sigma)$, where $x\in \left[l\right]^d$ labels the cell and $\sigma\in \Sigma$ labels the item within the cell.
The lattice Hamiltonian is of the form
\begin{equation}\label{eq:gen-H0}
    H_0 \ket{x,\sigma} = \sum\limits_{\substack{\delta\in\Delta\\ \sigma'\in\Sigma}} h_{\delta\sigma\sigma'}\ket{x+\delta,\sigma'}
\end{equation}
for some fixed finite set $\Delta\subset \mathbb Z^d$ with $-\Delta =\Delta$, and $h_{\delta\sigma\sigma'} = h_{-\delta\sigma'\sigma}^*$ to ensure that $H_0$ is Hermitian.
Translational invariance implies that $H_0$ is block diagonal in the Fourier basis \eqref{eq:grid-fourier}, i.e.\ (generalising \eqref{eq:grid-fourier-coeff1}),
\begin{equation}\label{eq:gen-blockdiagonal}
    H_0\ket{k,\sigma} = \sum\limits_{\substack{\delta\in\Delta\\ \sigma'\in\Sigma}}  h_{\delta\sigma\sigma'} e^{-ik\cdot\delta}\ket{k,\sigma'},
\end{equation}
such that \eqref{eq:grid-directsum} and \eqref{eq:grid-block1} hold. Diagonalising the $r\times r$ matrices $H_0(k)$ with matrix elements
\begin{equation}\label{eq:gen-blockelements}
    \left[H_0(k)\right]_{\sigma\sigma'} =  \sum_{\delta\in\Delta} h_{\delta\sigma\sigma'} e^{-ik\cdot\delta}
\end{equation} gives the dispersion relation $\E_i(k)$, with $i\in\left[r\right]$, for $H_0$.

\subsection{Lattice Hamiltonians with Dirac points} \label{sec:gen-dirac}

Diagonalising \eqref{eq:gen-blockelements} gives $r$ eigenvalues that can be collected into $r$ "energy bands" $\E_1(k),\ldots,\E_r(k)$.
We make the following assumptions about the dispersion relation of $H_0$.

\begin{enumerate}
  \item \label{en:gen1} $\E_1(k),\ldots,\E_m(k)$ have $D$ Dirac points at $k=\tilde k^{(1)},\ldots,\tilde k^{(D)}$ of the same energy for some $m\in\left[r\right]$. By an overall energy shift, we can assume without loss of generality that $\E_i(\tilde k^{(j)})= 0$ for all $i\in\left[m\right]$ and $j\in\left[D\right]$.
  \item \label{en:gen2} All other eigenvalues are nonzero away from the Dirac points.
  \item \label{en:gen3} $\E_{m+1},\ldots,\E_r$ are bounded away from zero.
  \item \label{en:gen4} There exists some $\tilde k\in\left[-\pi,\pi\right]^d$ such that, for all $j\in\left[D\right]$, the coordinates of  $\tilde k-\tilde k^{(j)}$ are rational multiples of $\pi$.
  \item \label{en:gen5} $\chi_\sigma^{(j)} := \|P_j \ket{\tilde k^{(j)},\sigma}\!\|^2\neq 0$ for all $\sigma\in\Sigma$, where $P_j$ is the projector onto the intersection of $\HH_{\tilde k^{(j)}}$ with the kernel of $H_0$ (i.e., onto the eigenstates corresponding to the energy bands $\E_i(\tilde k^{(j)})$ for $i \in [m]$).
\end{enumerate}

Assumption \ref{en:gen2} implies in particular that $\E_1,\ldots,\E_m$ cannot have common Dirac points at other energies.
Assumption \ref{en:gen3} ensures that the behaviour of the algorithm is dominated by the linear behaviour near the Dirac points.
Assumption \ref{en:gen4} is made for simplicity. If no such $\tilde k$ exists and some coordinates of $\tilde k^{(j)}$ are irrational multiples of $\pi$, suitable convergents of the prefactors can be considered instead.
Notice that since the matrix elements \eqref{eq:gen-blockelements} of $H_0(\tilde k^{(j)})$ are independent of $N$, so are the $\chi_\sigma^{(j)}$.
Assumption \ref{en:gen5} is trivially satisfied whenever $m=r$ (i.e., when all energy bands have a Dirac point, such as in Section~\ref{sec:grid}), since in this case $\chi_\sigma^{(j)}=1$.

Let the marked item be $\ket{w,\alpha}$. Define the normalised states
\begin{equation}
  \ket{s_\sigma^{(j)}} := \frac{1}{\sqrt{\chi_\sigma^{(j)}}} P_j\ket{\tilde k^{(j)},\sigma}.
\end{equation}
Take the starting state to be
\begin{equation}\label{eq:gen-start}
    \ket s := \frac{1}{\sqrt{\chi_\alpha}}\sum_{j=1}^D e^{-i \tilde k^{(j)}\cdot w}\sqrt{\chi_\alpha^{(j)}}\ket{s_\alpha^{(j)}},
\end{equation}
where
\begin{equation}
    \chi_\alpha := \sum_{j=1}^D \chi_\alpha^{(j)}.
\end{equation}

The state $\ket s$  depends on the unknown marked item via the relative phases of \eqref{eq:gen-start} and via $\alpha$. However, there are only $r$ possible values for $\alpha$ and, by assumption \ref{en:gen4}, the phases $e^{-i\tilde k^{(j)}\cdot w}=e^{i(\tilde k -\tilde k^{(j)})\cdot w}e^{-i\tilde k\cdot w}$ can only take some constant number of possible values, independent of $N$. Thus there are only $O(1)$ possible starting states for any given $\alpha$ (a trivial upper bound on this number is the least common multiple of all denominators of the rational numbers in assumption \ref{en:gen4}). Running the algorithm for every possible starting state only increases the running time by a constant factor.

We evolve $\ket s$ with the full Hamiltonian
\begin{equation}\label{eq:gen-H}
    H = H_0 + \Horacle,
\end{equation}
with $\Horacle$ specified below.

Define $F(E)$ as in \eqref{eq:grid-F-def}. One easily checks that \eqref{eq:grid-projection-norm} generalises to
\begin{equation}
    \left\|P_j\ket{w,\alpha}\right\|^2 = \frac{\chi_\alpha^{(j)}}{ n},
\end{equation}
so that \eqref{eq:grid-separate-0} and \eqref{eq:grid-separate-0-1} generalise to
\begin{align}
  F(E) &= -\frac{1}{E}\sum_{j=1}^D\left\|P_j\ket{w,\alpha}\right\|^2  + \bra{\tilde w}\left( H_0-E\right)^{-1} \ket{\tilde w} \nonumber \\
        &= -\frac{\chi_\alpha}{nE} + \bra{\tilde w}(\tilde H_0-E)^{-1} \ket{\tilde w},\label{eq:gen-separate-0-1}
\end{align}
where $\tilde H_0$ is the restriction of $H_0$ to the orthogonal complement of the kernel and as such is invertible.
Assuming that $|E|\ll |\E_i(k^{(j)})|$ for all $i\in\left[m\right]$, $j\in \left[D\right]$, and all $k\neq \tilde k^{(j)}$, we can Taylor expand the second term in \eqref{eq:gen-separate-0-1} as
\begin{align}\label{eq:gen-taylor1}
    F(E)&\approx -\frac{\chi_\alpha}{nE} + \bra{w,\alpha} \tilde H_0^{-1}\ket{w,\alpha} \nonumber \\
        &\quad +  E \bra{w,\alpha} \tilde H_0^{-2}\ket{w,\alpha}.
\end{align}
This can be justified in a similar fashion to Section~\ref{sec:grid-rigorous}.
By assumptions \ref{en:gen2} and \ref{en:gen3}, the behaviour of the last two terms for $n\rightarrow\infty$ is dominated by the behaviour around $k=\tilde k^{(j)}$. In particular, linearity of the dispersion relation around the Dirac points ensures that $\bra{w,\alpha} \tilde H_0^{-1}\ket{w,\alpha}$ is bounded and in fact converges to some value $I_{1,d}$, while $\bra{w,\alpha} \tilde H_0^{-2}\ket{w,\alpha}$ converges to some value $I_{2,d}$ for $d>2$ and diverges logarithmically for $d=2$. Thus, we can expand these expressions in the eigenbasis of $H_0$ and approximate them by integrals (similarly to \eqref{eq:grid-integral}) so that
\begin{equation}\label{eq:gen-taylor2}
    F(E) \approx -\frac{\chi_\alpha}{nE} + I_{1,d} + I_{2,d}E.
\end{equation}
To find an eigenvalue gap of $O(1/\sqrt N)$ (or $O(1/\sqrt{N\log N})$ in $d=2$), the eigenvalue condition should be $F(E)=I_{1,d}$; we choose $\Horacle$ to achieve this. The choice of $\Horacle$ thus depends on the value of $I_{1,d}$. In particular, we choose qualitatively different oracle terms depending on whether $I_{1,d}$ is zero.

\begin{icase}
Suppose that $I_{1,d}=0$. In particular, this holds whenever $H_0(k)$ is $0$ on the diagonal (or equivalently, whenever the lattice is $r$-partite with the vertices partitioned according to the value of $\sigma \in \Sigma$) and the dispersion relation splits into two (possibly degenerate) energy bands that are symmetric with respect to the Dirac point,
as in our example in Section~\ref{sec:grid}. In this case we can choose the oracle Hamiltonian to be
\begin{equation}\label{eq:gen-oracle1}
    \Horacle = - H_0\ket{w,\alpha}\bra{w,\alpha} - \ket{w,\alpha}\bra{w,\alpha} H_0.
\end{equation}
\end{icase}

\begin{icase}
Suppose that $I_{1,d}\neq 0$. This was the case in \cite{CG03}, and as in those algorithms, we can choose the oracle Hamiltonian to be
\begin{equation}\label{eq:gen-oracle2}
    \Horacle = -\frac{1}{I_{1,d}} \ket{w,\alpha}\bra{w,\alpha}.
\end{equation}
The prefactor of $1/I_{1,d}$ plays the role of the parameter $\gamma$ discussed in Section~\ref{sec:grid-finetuning}.
If $\ket{\psi_a}$ is an eigenvector of $H$ with eigenvalue $E_a$ that is not in the spectrum of $H_0$, then $H\ket{\psi_a}=E_a\ket{\psi_a}$ is equivalent to
\begin{equation}
    I_{1,d}\ket{\psi_a} = (H_0-E_a)^{-1} \ket{w,\alpha}\braket{w,\alpha|\psi_a},
\end{equation}
so that our eigenvalue condition on $E_a$ is
\begin{equation}\label{eq:gen-condition2}
    F(E_a)=I_{1,d},
\end{equation}
as required.
\end{icase}

In both cases, we obtain two approximate eigenvalues
\begin{equation}
    E_\pm \approx \pm\sqrt{\frac{\chi_\alpha}{nI_{2,d}}}.
\end{equation}
The overlaps of the corresponding eigenvectors with $\ket{s_\alpha^{(j)}}$ can be calculated similarly to \eqref{eq:grid-spsi}, and are given by
\begin{align}\label{eq:gen-overlap-j}
    \braket{\psi_\pm|s_\alpha^{(j)}} &= -\frac{e^{i\tilde k^{(j)}\cdot w}}{E_\pm }\sqrt{\frac{\chi_\alpha^{(j)}}{nF'(E_\pm)}} \\
        &\approx \mp e^{i\tilde k^{(j)}\cdot w}\sqrt{\frac{\chi_\alpha^{(j)}}{2\chi_\alpha}},
\end{align}
so that from \eqref{eq:gen-start}, $\braket{\psi_\pm|s} \approx \mp\frac1{\sqrt 2}$.
This ensures that $\ket s$ is supported essentially only on the two-dimensional subspace spanned by $\ket{\psi_\pm}$.

Finally, the same calculations as in Section~\ref{sec:grid-analysis} show that for $d>2$, evolving $\ket s$ for a time $T=O(\sqrt N)$ results in a constant overlap with $H_0\ket{w,\alpha}/\sqrt{\bra{w,\alpha} H_0^2 \ket{w,\alpha}}$ in case 1 and $\ket{w,\alpha}$ in case 2, while for $d=2$, evolving for $T=O(\sqrt{N\log N})$ results in an overlap of $\Omega(1/\sqrt{\log N})$.

\begin{figure}[h]
    \centering
    \subfloat[][Disconnected]{
      \begin{tikzpicture}[scale=0.6]
        \clip (-2.5,-2.5) rectangle (2.5,2.5);
          \foreach \i in {-3,-2,...,3}
          {
            \foreach \j in {-3,-2,...,3}
            {
                \ifthenelse{\cnttest{\i}={0} \AND \cnttest{\j}={0} }
                {
                    \draw (\i,\j) node [largeorangedot] {};
                }
                {
                    \ifthenelse{ \( \cnttest{\i}={0} \AND \cnttest{\j*\j}={1} \) \OR \( \cnttest{\j}={0} \AND \cnttest{\i*\i}={1} \)}
                    {
                    }
                    {
                         \draw[greydot] (\i-1,\j) node [greydot] {}
                            -- (\i,\j)
                            -- (\i+1,\j) node [greydot] {}
                            ;
                         \draw[greydot] (\i,\j-1) node [greydot] {}
                            -- (\i,\j) node [greydot] {}
                            -- (\i,\j+1) node [greydot] {}
                            ;
                    }
                }
                \node[bluedot] (u) at (0,1) {};
                \node[bluedot] (d) at (0,-1) {};
                \node[bluedot] (l) at (-1,0) {};
                \node[bluedot] (r) at (1,0) {};
                \path (r) edge [blue,loop left, in=150,out=210,looseness=10] node {} (r);
                \path (l) edge [blue,loop left, in=30,out=-30,looseness=10] node {} (l);
                \path (u) edge [blue,loop left, in=-60,out=-120,looseness=10] node {} (u);
                \path (d) edge [blue,loop left, in=60,out=120,looseness=10] node {} (d);
            }
          }
    \end{tikzpicture}
    \label{subfig:oracles-1}}
    \qquad
    \subfloat[][On-site potential]{
      \begin{tikzpicture}[scale=0.6]
        \clip (-2.5,-2.5) rectangle (2.5,2.5);
          \foreach \i in {-3,-2,...,3}
          {
            \foreach \j in {-3,-2,...,3}
            {
                \ifthenelse{\cnttest{\i}={0} \AND \cnttest{\j}={0} }
                {

                }
                {
                         \draw[greydot] (\i-1,\j) node [greydot] {}
                            -- (\i,\j)
                            -- (\i+1,\j) node [greydot] {}
                            ;
                         \draw[greydot] (\i,\j-1) node [greydot] {}
                            -- (\i,\j) node [greydot] {}
                            -- (\i,\j+1) node [greydot] {}
                            ;
                }
                \node[largeorangedot] (0) at (0,0) {};
                \path (0) edge [MarkedOrange,loop left, in=15,out=75,looseness=10] node {} (0);
            }
          }
    \end{tikzpicture}
    \label{subfig:oracles-2}}
    \caption{Effects of the different choices of $\Horacle$. (a) The choice in \eqref{eq:gen-oracle1} disconnects the marked vertex from the rest of the lattice and creates  on-site potentials at the neighbours of the marked item.
    (b) The choice in \eqref{eq:gen-oracle2} creates an on-site potential at the marked item.
    }
    \label{fig:gen-oracles}
\end{figure}
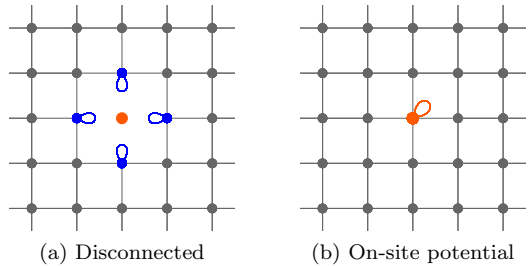

We briefly interpret the two different choices of $\Horacle$.
In the first case, \eqref{eq:gen-oracle1} modifies the strength of the transition amplitudes between the marked item and its neighbours. Specifically, \eqref{eq:gen-oracle1} modifies the Hamiltonian such that $H\ket{w,\alpha}=0$ and $\bra{w'}H\ket{w'}\neq 0$ for all neighbours $\ket{w'}$ of $\ket{w,\alpha}$. The first condition implies that the probability amplitude on the marked item is invariant under evolution with $H$, so the marked item is disconnected from the rest of the lattice. The latter condition creates "on-site potentials" at the states $\ket{w'}$, giving loops in the graph structure (see Fig.~\ref{subfig:oracles-1}).
On the other hand, \eqref{eq:gen-oracle2} creates an on-site potential directly at the marked item
(see Fig.~\ref{subfig:oracles-2}).
Other possible oracle Hamiltonians involving single-edge alterations and  additional vertices are briefly mentioned in \cite{FGT13}.

\subsection{Examples}\label{sec:examples}

\begin{example}
  To recover the example of Section~\ref{sec:grid}, we set $r=2^d$, $\Sigma=\mathbb Z_2^d$, and $\Delta = \{\pm e_i \colon i\in\left[d\right]\}$. By comparing \eqref{eq:grid-H0-position} with \eqref{eq:gen-H0} (or, equivalently \eqref{eq:grid-fourier-coeff1} with \eqref{eq:gen-blockdiagonal}), we can read off the coefficients as
  \begin{align}
    h_{0,\sigma,\sigma+e_i} &= (-1)^{\sigma_1+\cdots+\sigma_{i-1}},
    \label{eq:ex-grid-0}
    \\
    h_{e_i,\sigma,\sigma+e_i} &=
     \begin{cases}
        0, & \sigma_i = 0,\\
        -(-1)^{\sigma_1+\cdots+\sigma_{i-1}}, & \sigma_i = 1,
     \end{cases} \label{eq:ex-grid-pos} \\
    h_{-e_i,\sigma,\sigma+e_i} &=
    \begin{cases}
        -(-1)^{\sigma_1+\cdots+\sigma_{i-1}}, & \sigma_i = 0,\\
        0, & \sigma_i = 1,
     \end{cases}
     \label{eq:ex-grid-neg}
  \end{align}
  with all other coefficients vanishing.
  Considering the nonzero coefficients, we see that the underlying graph is simply a hypercubic lattice.

  \begin{figure}[h]
    \centering
     \begin{tikzpicture}[scale=0.6]
        \clip (-3.5,-3) rectangle (4.5,3);
        \draw (-3,0) -- (-2,0);
        \foreach \i in {-1,0,1}
        {
            \filldraw[dottedcell] (2*\i-0.3,-0.5) rectangle (2*\i+1.3,0.5);
            \draw  (2*\i,0) node [reddot] {}
                -- (2*\i+1,0) node [bluedot] {}
                -- (2*\i+2,0);
            \node[above,color=red,rotate=45] at (2*\i+0.8,1) {\footnotesize $\sigma_i=0$};
            \node[above,color=blue,rotate=45] at (2*\i + 1 + 0.8,1) {\footnotesize $\sigma_i=1$};
        }
        \node[below] at (-1.5,-0.5) {$x-e_i$};
        \node[below] at (0.5,-0.6) {$x$};
        \node[below] at (2.5,-0.5) {$x+e_i$};
        \draw[->,very thick] (0,-2) -- (1,-2);
        \node[right] at (1,-2) {$e_i$};
     \end{tikzpicture}
     \caption{Recovering the edges of the graph from the coefficients \eqref{eq:ex-grid-0}--\eqref{eq:ex-grid-neg}. For each nonzero $h_{\delta,\sigma,\sigma'}$, there is an edge from $\ket{x,\sigma}$ to $\ket{x+\delta, \sigma'}$.}
     \label{fig:ex-grid}
  \end{figure}
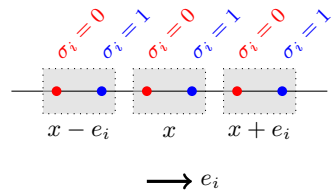

  Figure \ref{fig:ex-grid} depicts the edges corresponding to nonzero coefficients.
  Equation \eqref{eq:ex-grid-0} implies that there is an edge in any given direction $i$ within a given cell.
  Equation \eqref{eq:ex-grid-pos} implies that there is an edge from the $\sigma_i=1$ (right) vertices of a given cell $x$ to the $\sigma_i=0$ (left) vertices of the cell $x+e_i$, and similarly \eqref{eq:ex-grid-neg} implies that there is an edge from the $\sigma_i=0$ (left) vertices of a given cell $x$ to the $\sigma_i=1$ (right) vertices of the cell $x-e_i$.
  Repeating this procedure for all directions $i \in [d]$, we see that the graph structure of a $d$-dimensional hypercubic lattice is recovered (see Fig.~\ref{fig:grid-crystal}).
  With the coefficients given by \eqref{eq:ex-grid-0}--\eqref{eq:ex-grid-neg}, the eigenvalues of the $2^d \times 2^d$ matrices \eqref{eq:gen-blockelements} are given by \eqref{eq:grid-dispersion}.
\end{example}

\begin{example}[Honeycomb lattice]
  The best known example of a lattice with Dirac points may be the honeycomb lattice in $d=2$, the lattice structure of graphene. We can recover this lattice in our formalism by setting $r=2$
  and taking
  \begin{equation}\label{eq:ex-graphene-H0}
    H_0(k) = \left(
        \begin{array}{cc}
          0 & h(k)^*  \\
          h(k) & 0
        \end{array}
    \right),
  \end{equation}
  where $h(k):=1+e^{-ik_x}+e^{-i(k_x+k_y)}$.
  It is easy to see that, with this choice, $H_0$ is the adjacency matrix of a graph that is isomorphic to the standard honeycomb lattice (see Fig.~\ref{fig:graphene}).

  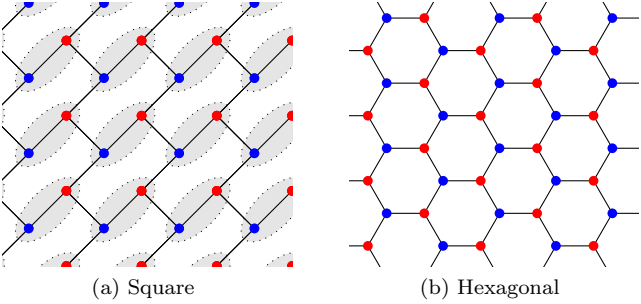
\begin{figure}[h]
    \centering
    \subfloat[][Square]{
      \begin{tikzpicture}[scale=0.5]
        \clip (-3.7,-4) rectangle (4,3);
          \foreach \i in {-3,-2,...,3}
          {
            \foreach \j in {-3,-2,...,3}
            {
                \filldraw[dottedcell, rotate around={45:(2*\i-0.5,2*\j-0.5)}] (2*\i-0.5,2*\j-0.5) ellipse (1.1 and 0.5);
        	   \draw  (2*\i,2*\j) -- (2*\i+1,2*\j+1);
               \draw(2*\i-2,2*\j) node [reddot] {} -- (2*\i-1,2*\j-1) -- (2*\i-2,2*\j-2) node [reddot] {};
        	   \draw  (2*\i,2*\j) -- (2*\i+1,2*\j-1) node [bluedot]{} ;
               \draw (2*\i,2*\j) node [reddot] {} -- (2*\i-1,2*\j-1) node [bluedot]{};
            }
          }
    \end{tikzpicture}}
    \qquad
    \subfloat[][Hexagonal]{
      \begin{tikzpicture}[scale=0.5]
        \clip (-3.5,-4) rectangle (4.2,3);
          \foreach \i in {-4,-3,...,4}
          {
            \foreach \j in {-4,-3,...,4}
            {
                \foreach \a in {0,120,-120}
                {
                    \draw (cos{60}+cos{60}*\i+\i + cos{60}*\j+\j,sin{60}+sin{60}*\i-sin{60}*\j) -- +(\a:1);
                }
                \node[reddot] at (cos{60}*\i+\i + cos{60}*\j+\j,sin{60}*\i-sin{60}*\j) {};
                \node[bluedot] at (cos{60}+cos{60}*\i+\i + cos{60}*\j+\j,sin{60}+sin{60}*\i-sin{60}*\j) {};
            }
          }
    \end{tikzpicture}}
    \caption{Drawings of honeycomb lattices. (a) Bipartite square lattice with two items per cell as in \eqref{eq:ex-graphene-H0}. (b) Standard drawing of a honeycomb lattice.}
    \label{fig:graphene}
  \end{figure}

  The dispersion relation of this Hamiltonian is
  \begin{align}\label{eq:ex-graphene-dispersion}
    \E(k) &= \pm |{h(k)}| \\
          &= \pm \sqrt{3 + 2 (\cos k_x+ \cos k_y + \cos(k_x+k_y))},
  \end{align}
  which has two Dirac points at $k_x=k_y=\pm 2\pi/3$. The special case of spatial search on the honeycomb lattice is studied  in detail in \cite{FGT13}.
\end{example}

\begin{example}[Kagome lattice]
  Another example in $d=2$ is given by the adjacency matrix of a Kagome lattice (see Fig.~\ref{fig:kagome}). We can recover this lattice in our formalism by setting $r=2$ and taking
  \begin{equation}\label{eq:ex-kagome-H0}
    H_0(k) = \left(
        \begin{array}{ccc}
          -1 & g(k_y) & g(-k_x+k_y) \\
          g(-k_y) & -1 & g(-k_x) \\
          g(k_x-k_y) & g(k_x) & -1
        \end{array}
    \right),
  \end{equation}
  where $g(k):= 1+e^{ik}$. The diagonal elements only provide an overall energy shift and are included for convenience.

  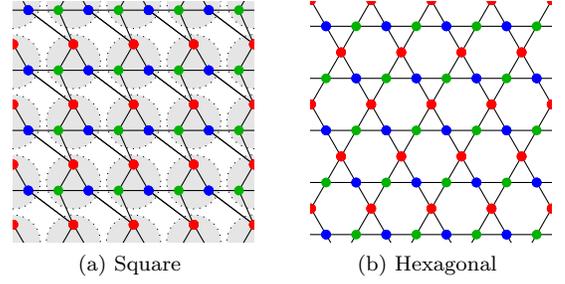
\begin{figure}[h]
    \subfloat[][Square]{
          \begin{tikzpicture}[scale=0.4]
                        \clip (-4,-4) rectangle (4,4);
                        \foreach \i in {-3,-2,...,3}
                        {
                            \foreach \j in {-3,-2,...,3}
                            {
                                \filldraw[dottedcell] (2*\i,2*\j) circle (0.9);
                                \draw (2*\i+0.5,2*\j-sin{60}*1/3)
                                   -- (2*\i+0.5-2,2*\j-sin{60}*1/3) node[bluedot] {};
                                \draw  (2*\i+2,2*\j-2+2*sin{60}/3) node[reddot] {}
                                    -- (2*\i+0.5,2*\j-sin{60}*1/3) node[bluedot] {}
                                    -- (2*\i,2*\j+2*sin{60}/3) node[reddot] {}
                                    -- (2*\i-0.5,2*\j-sin{60}*1/3) node[greendot] {}
                                    -- (2*\i,2*\j-2+2*sin{60}/3) node[reddot] {};
                                \draw  (2*\i,2*\j+2*sin{60}/3) node[reddot] {}
                                    -- (2*\i-2+0.5,2*\j+2-sin{60}*1/3) node[bluedot] {};
                            }
                        }
        \end{tikzpicture}}
    \qquad
    \subfloat[][Hexagonal]{
        \begin{tikzpicture}[scale=0.4]
                        \clip (-4,-4) rectangle (4,4);
                        \foreach \i in {-3,-2,...,3}
                        {
                            \foreach \j in {-3,-2,...,3}
                            {
                                \draw (2*\i+2*\j*cos{60},2*\j*sin{60}+2*sin{60}/3)
                                   -- (2*\i+2*\j*cos{60} - 2*cos{60},2*\j*sin{60}+2*sin{60}/3 - 2*sin{60})  node[reddot] {};
                                \draw (2*\i+2*\j*cos{60},2*\j*sin{60}+2*sin{60}/3)
                                   -- (2*\i+2*\j*cos{60} + 2*cos{60},2*\j*sin{60}+2*sin{60}/3 - 2*sin{60}) node[reddot] {};
                                \draw (2*\i+2*\j*cos{60}+0.5,2*\j*sin{60}-sin{60}*1/3)
                                   -- (2*\i+2*\j*cos{60}+0.5-2,2*\j*sin{60}-sin{60}*1/3) node[bluedot] {};
                                \draw(2*\i+2*\j*cos{60},2*\j*sin{60}+2*sin{60}/3) node[reddot] {};
                                \draw(2*\i+2*\j*cos{60}+0.5,2*\j*sin{60}-sin{60}*1/3) node[bluedot] {};
                                \draw(2*\i+2*\j*cos{60}-0.5,2*\j*sin{60}-sin{60}*1/3) node[greendot] {};
                            }
                        }
        \end{tikzpicture}}
    \caption{Drawings of Kagome lattices. (a) Kagome lattice as tripartite square lattice with three items per cell using \eqref{eq:ex-kagome-H0}. (b) Standard drawing of a Kagome lattice.}
    \label{fig:kagome}
  \end{figure}

  The dispersion relation of this Hamiltonian comprises three energy bands given by
  \begin{align}
    \E_{\pm} (k) &= \pm \sqrt{3 + 2(\cos k_x+ \cos k_y  +  \cos(k_x - k_y))}, \\
    \E_3(k) &= -3.
  \end{align}
  The top two bands $\E_{\pm}$ have two Dirac points at $k_x=-k_y = \pm 2\pi/3$ of energy $\E=0$. Notice that $\E_3$ is bounded away from $0$ (since it is constant) and it is easy to verify that all the assumptions of Section~\ref{sec:gen-dirac} are satisfied. Unlike the previous examples, $I_{1,2}\approx -4.39 \neq 0$, so the oracle Hamiltonian can be chosen as in \eqref{eq:gen-oracle2}.
\end{example}

\begin{example}\label{ex:dirac}
 Again in $d=2$ and $r=2$, we can consider the Hamiltonian
 \begin{equation} \label{eq:ex-dirac-H0}
     H_0(k) = \left(
        \begin{array}{cc}
          \gamma c(k)& \omega s(k)^*\\
         \omega s(k) & -\gamma c(k)
        \end{array}
    \right),
 \end{equation}
 where $s(k) := \sin k_x - i\sin k_y$, $c(k) := 2-\cos k_x - \cos k_y$, and $\gamma,\omega\in\mathbb R$. This reproduces the same dispersion relation found in \cite{CG04},
 \begin{equation}
    \E(k) = \pm \sqrt{ \omega^2\lvert s(k)\rvert^2 + \gamma^2\lvert c(k)\rvert^2}.
 \end{equation}
 As such, the choice \eqref{eq:ex-dirac-H0} effectively embeds the additional "spin" degrees of freedom introduced in \cite{CG04} into the lattice as additional vertices. A similar approach also recovers the Hamiltonian from \cite{CG04} in higher dimensions.
 The diagonal terms in \eqref{eq:ex-dirac-H0} ensure the uniqueness of the Dirac point at $k_x=k_y=0$. However, using the results of Section~\ref{sec:gen-dirac}, we can obtain a working algorithm even when $\gamma=0$. In this case, the underlying graph is not only bipartite but also disconnected (see Fig.~\ref{fig:ex-dirac}). The connected components are both isomorphic to two-dimensional square lattices and the underlying Hamiltonian acts on these as
 \begin{align}
    H_0\ket v &= i(-1)^y(\ket{v+e_x} - \ket{v-e_x}) \nonumber \\
    &\quad + (-1)^y (\ket{v+e_y} - \ket{v-e_y}). \label{eq:ex-dirac-H0-new}
 \end{align}
 This gives an alternative Hamiltonian for searching a two-dimensional square lattice with near-quadratic speedup.

 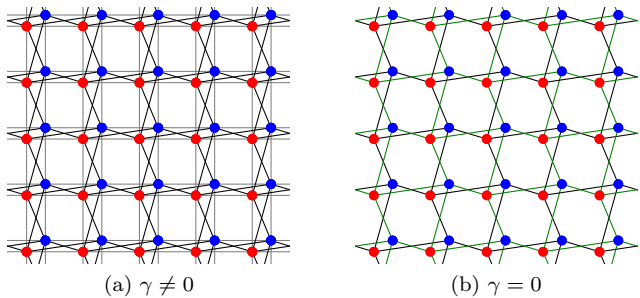
\begin{figure}[h]
   \centering
   \subfloat[][$\gamma\neq 0$]{
     \begin{tikzpicture}[scale=0.5]
     \clip (-3.5,-3.3) rectangle (4,3.5);
     \foreach \i in {-2,-1,...,2}
      {
        \foreach \j in {-2,-1,...,2}
        {
          \draw[color=gray] (3*\i,3*\j) node [reddot] {}
           -- (3*\i,3*\j+1.5) node [reddot] {}
           -- (3*\i+1.5,3*\j+1.5) node [reddot] {}
           -- (3*\i+1.5,3*\j) node [reddot] {}
           -- (3*\i,3*\j) node [reddot] {}
           -- (3*\i-1.5,3*\j) node [reddot] {}
           -- (3*\i-1.5,3*\j-1.5) node [reddot] {}
           -- (3*\i,3*\j-1.5) node [reddot] {}
           -- (3*\i,3*\j) node [reddot] {}
           ;
           \draw[color=gray] (3*\i+0.5,3*\j+0.3) node [bluedot] {}
           -- (3*\i+0.5,3*\j+1.5+0.3) node [bluedot] {}
           -- (3*\i+1.5+0.5,3*\j+1.5+0.3) node [bluedot] {}
           -- (3*\i+1.5+0.5,3*\j+0.3) node [bluedot] {}
           -- (3*\i+0.5,3*\j+0.3) node [bluedot] {}
           -- (3*\i-1.5+0.5,3*\j+0.3) node [bluedot] {}
           -- (3*\i-1.5+0.5,3*\j-1.5+0.3) node [bluedot] {}
           -- (3*\i+0.5,3*\j-1.5+0.3) node [bluedot] {}
           -- (3*\i+0.5,3*\j+0.3) node [bluedot] {}
           ;
           \draw[color=black] (3*\i,3*\j) node [reddot] {}
           -- (3*\i+0.5,3*\j+1.5+0.3) node [bluedot]{}
           -- (3*\i+1.5,3*\j+1.5) node [reddot]{}
           -- (3*\i+1.5+0.5,3*\j+0.3) node [bluedot]{}
           -- (3*\i,3*\j) node [reddot]{}
           -- (3*\i-1.5+0.5,3*\j+0.3) node [bluedot]{}
           -- (3*\i-1.5,3*\j-1.5) node [reddot]{}
           -- (3*\i+0.5,3*\j-1.5+0.3) node [bluedot]{}
           -- (3*\i,3*\j)  node [reddot]{}
           ;
           \draw[color=black] (3*\i+0.5,3*\j+0.3) node [bluedot] {}
           -- (3*\i,3*\j+1.5) node [reddot]{}
           -- (3*\i+1.5+0.5,3*\j+1.5+0.3) node [bluedot]{}
           -- (3*\i+1.5,3*\j) node [reddot]{}
           -- (3*\i+0.5,3*\j+0.3) node [bluedot]{}
           -- (3*\i-1.5,3*\j) node [reddot]{}
           -- (3*\i-1.5+0.5,3*\j-1.5+0.3) node [bluedot]{}
           -- (3*\i,3*\j-1.5) node [reddot]{}
           -- (3*\i+0.5,3*\j+0.3)  node [bluedot]{}
           ;
        }
      }
    \end{tikzpicture}
    \label{subfig:dirac1}}
   \qquad
   \subfloat[][$\gamma=0$]{
     \begin{tikzpicture}[scale=0.5]
     \clip (-3.5,-3.3) rectangle (4,3.5);
     \foreach \i in {-2,-1,...,2}
      {
        \foreach \j in {-2,-1,...,2}
        {
           \draw[color=black] (3*\i,3*\j) node [reddot] {}
           -- (3*\i+0.5,3*\j+1.5+0.3) node [bluedot]{}
           -- (3*\i+1.5,3*\j+1.5) node [reddot]{}
           -- (3*\i+1.5+0.5,3*\j+0.3) node [bluedot]{}
           -- (3*\i,3*\j) node [reddot]{}
           -- (3*\i-1.5+0.5,3*\j+0.3) node [bluedot]{}
           -- (3*\i-1.5,3*\j-1.5) node [reddot]{}
           -- (3*\i+0.5,3*\j-1.5+0.3) node [bluedot]{}
           -- (3*\i,3*\j)  node [reddot]{}
           ;
           \draw[color=green!50!black] (3*\i+0.5,3*\j+0.3) node [bluedot] {}
           -- (3*\i,3*\j+1.5) node [reddot]{}
           -- (3*\i+1.5+0.5,3*\j+1.5+0.3) node [bluedot]{}
           -- (3*\i+1.5,3*\j) node [reddot]{}
           -- (3*\i+0.5,3*\j+0.3) node [bluedot]{}
           -- (3*\i-1.5,3*\j) node [reddot]{}
           -- (3*\i-1.5+0.5,3*\j-1.5+0.3) node [bluedot]{}
           -- (3*\i,3*\j-1.5) node [reddot]{}
           -- (3*\i+0.5,3*\j+0.3)  node [bluedot]{}
           ;
        }
      }
    \end{tikzpicture}
    \label{subfig:dirac2}}
   \caption{(a) Graph induced by \eqref{eq:ex-dirac-H0} for generic values of $\gamma$. (b) If $\gamma=0$, the graph is both bipartite and disconnected and the two connected components are both isomorphic to a two-dimensional square lattice.}
   \label{fig:ex-dirac}
 \end{figure}

 Notice, however, that although each component only contains one vertex from each cell, the Hamiltonian \eqref{eq:ex-dirac-H0-new} is nonhomogenous in the $y$ direction, so we must combine the vertices into new cells of size $r=2$. Both \eqref{eq:ex-dirac-H0-new} and \eqref{eq:grid-H0} are defined on the same underlying lattice and give algorithms with the same complexity, but they are inequivalent Hamiltonians. Indeed, the two Hamiltonians have different symmetries as \eqref{eq:ex-dirac-H0-new} is uniform in the $x$ direction, resulting in $r=2$, whereas \eqref{eq:grid-H0} has four items per cell. Furthermore, the dispersion relation of  \eqref{eq:grid-H0} has a unique Dirac point, whereas \eqref{eq:ex-dirac-H0-new} has two.
\end{example}

\section{Discussion}
\label{sec:discussion}

We presented a general framework for describing spatial search algorithms using continuous-time quantum walks. Using the linearity of the dispersion relation near Dirac points, we constructed search algorithms that provide the optimal quantum speedup of $O(\sqrt N)$ in $d>2$ dimensions and have complexity $O(\sqrt N \log N)$ in $d=2$. In particular, we constructed such algorithms for hypercubic lattices in $d\geq 2$ dimensions.

The algorithms presented here are closely related to the ones described in \cite{CG04} and generalise the results from \cite{FGT13}. Inspired by the Dirac equation, \cite{CG04} introduced additional "spin" degrees of freedom, similar to "coin" registers for discrete-time walks, to obtain a Hamiltonian exhibiting a Dirac point. Our framework can be used to construct equivalent Hamiltonians without external memory by embedding these additional degrees of freedom into the lattice as additional vertices. The naive way of doing this introduces many additional nonzero transition amplitudes (i.e., edges) in $H_0$ so that the underlying graph is not isomorphic to a hypercubic lattice. However, with further modifications as presented in Sections~\ref{sec:grid} and \ref{sec:examples} (Example~\ref{ex:dirac}), we showed it is possible to recover the structure of a hypercubic lattice.

In high dimensions, the algorithm presented in Section~\ref{sec:grid} requires large cells of size $2^d$. The results from \cite{CG04} show that this can be reduced to $d+1$. However, unlike in \cite{CG04}, those spin registers do not manifest themselves as additional memory in our algorithm: every vertex corresponds to a distinct possible marked item. The procedure is versatile and can in principle be applied to any continuous-time quantum walk spatial search algorithm to reduce the external memory at the cost of possibly introducing additional edges into the graph and requiring multiple runs to ensure success.

Note that the actual complexity of the spatial search problem in $d=2$ is still an open question. Tulsi \cite{Tul08} proposed a method for reducing the complexity from $O(\sqrt N \log N)$ to $O(\sqrt {N \log N})$ for constant probability of success by controlling the walk using an ancilla qubit. It would be interesting to improve the complexity further or to show that no such improvement is possible.

We remark that if the locality constraint is relaxed to only require an interaction strength that monotonically decreases with the distance, it is possible to construct a Hamiltonian that achieves the optimal $O(\sqrt N)$ running time in $d=2$. Specifically, it suffices to choose $\bra{v}H_0\ket{v'} \approx \dd(x,y)^{-(2-\epsilon)}$  decaying as an almost quadratic power law (for any $\epsilon>0$). However, such a power-law decay should not be considered local.

\acknowledgments

Y.G. thanks Denis Dalidovich, Wen Wei Ho, and Heidar Moradi for helpful discussions.
This work was supported in part by NSERC, the Ontario Ministry of Research and Innovation, and the US ARO.
Research at Perimeter Institute is supported by the Government of Canada through Industry Canada and by the Province of Ontario through the Ministry of Research and Innovation.


\end{document}